\documentclass[twocolumn,preprintnumbers,amsmath,amssymb,superscriptaddress]{revtex4}
\usepackage{graphicx}
\usepackage{dcolumn}
\usepackage{bm}
\usepackage{soul}
\usepackage{color}
\usepackage{epstopdf}
\usepackage[version=3]{mhchem}
\usepackage{lipsum}
\usepackage[outercaption]{sidecap}
\usepackage{floatrow}
\usepackage{hyperref}
\usepackage{multirow}
\begin{document}


\title{A Multiscale Approach to Structural Relaxation and Diffusion in Metallic Glasses}

\author{Anh D. Phan}
\email{anh.phanduc@phenikaa-uni.edu.vn}
\affiliation{Faculty of Materials Science and Engineering, Phenikaa University, Hanoi 12116, Vietnam}
\affiliation{Phenikaa Institute for Advanced Study, Phenikaa University, Hanoi 12116, Vietnam}
\author{Do T. Nga}
\affiliation{Institute of Physics, Vietnam Academy of Science and Technology, 10 Dao Tan, Ba Dinh, Hanoi 12116, Vietnam}
\author{Ngo T. Que}
\affiliation{Phenikaa Institute for Advanced Study, Phenikaa University, Hanoi 12116, Vietnam}
\author{Hailong Peng}
\affiliation{School of Materials Science and Engineering, Central South University, Changsha 410083, People’s Republic of China}

\author{Thongchanh Norhourmour}
\affiliation{Hanoi Pedagogical University 2, Nguyen Van Linh Street, Vinh Phuc, Vietnam}
\author{Le M. Tu}
\affiliation{Faculty of Materials Science and Engineering, Phenikaa University, Hanoi 12116, Vietnam}
\date{\today}

\begin{abstract}
Metallic glasses are promising materials with unique mechanical and thermal properties, but their atomic-scale dynamics remain challenging to understand. In this work, we develop a unified approach to investigate the glass transition and structural relaxation in CoCrNi, \ce{Zr_{46}Cu_{46}Al_8}, \ce{Zr_{50}Cu_{40}Al_{10}}, and \ce{Zr_{64.13}Cu_{15.75}Ni_{10.12}Al_{10}} metallic glasses. Molecular dynamics (MD) simulation is employed to analyze the radial distribution function at different temperatures and accurately determine the glass transition temperature. We then combine this temperature with the Elastically Collective Nonlinear Langevin Equation (ECNLE) theory to predict the temperature dependence of the structural relaxation time, $\tau_\alpha(T)$. By connecting $\tau_\alpha(T)$ to the diffusion constant, the ECNLE predictions of $\tau_\alpha(T)$ can be compared with those calculated from MD simulations or estimated based on the diffusion constant. By combining atomistic simulation with force-level statistical mechanics, our multiscale approach offers deeper insights into relaxation dynamics and diffusion across various timescales. The relationship between the glass transition and the liquidus temperature is elucidated. This study enhances understanding of the glassy dynamics and properties in complex amorphous materials. 
\end{abstract}

\maketitle
\section{Introduction}
Metallic glasses have gained significant attention as potential candidates for high-performance engineering applications due to their unique and remarkable mechanical properties \cite{26,27,28,29}. These materials are conventionally produced through rapid cooling (in the order of millions of degree Celsius per second), which prevents the formation of a crystalline structure and instead results in a fully disordered, glass-like atomic arrangement. Despite their amorphous structure, metallic glasses retain the electrical conductivity characteristic of metals. Understanding the structural relaxation dynamics of these amorphous materials is crucial for a range of applications, as it provides insight into how atomic-scale rearrangements occur within the complex free-energy landscape. This knowledge is essential for predicting and optimizing the performance and stability of metallic glasses across various engineering and technological fields.

The structural relaxation time is strongly affected by temperature, external pressure, and confinement conditions \cite{26,34,35,36}. At high temperatures, atoms are highly mobile and materials have a liquid-like character. As the temperature decreases, however, the dynamics slow down considerably and this cooling leads to a sharp increase in $\tau_\alpha$. This slowdown facilitates the formation of solid materials with remarkable mechanical properties including enhanced hardness and ductility. Around the glass transition temperature, $T_g$, thermodynamic and physical properties, such as viscosity and diffusivity, undergo abrupt changes, increasing by several orders of magnitude \cite{34,36}.

Theoretical and simulation approaches play a pivotal role in investigating key properties and the dynamics of metallic glasses. Molecular dynamics simulations have been widely used to compute the temperature-dependent diffusion constant and structural relaxation time by tracking atomic trajectories and correlating them with time \cite{3}. MD allows for detailed atomic-scale insights into the dynamic processes, but its accuracy is limited by the short timescales and finite system sizes accessible to simulations \cite{3,5}. The Vogel-Fulcher-Tammann (VFT) model is commonly employed to fit diffusion and relaxation data, estimate $T_g$, and extrapolate long-time behaviors. However, it is a phenomenological model that may oversimplify the complex structural dynamics in glasses. Meanwhile, the Elastically Collective Nonlinear Langevin Equation theory \cite{1,2,3,4,5,6,7,8,9,30,39,40,41,42} has offered a rigorous theoretical framework that captures the local and cooperative nature of structural relaxation and provides quantitative predictions for the temperature and pressure dependence of diffusion constant and relaxation time. ECNLE calculations have agreed well with experimental data and simulations of many metallic glasses, polymers, thermal liquids, and amorphous drugs \cite{1,2,3,4,5,6,7,8,9,30,39,40,41,42}. It can be challenging to apply the ECNLE theory to metallic glasses without knowing their glass transition temperature. 

This raises several questions: (1) How can ECNLE theory be effectively applied to metallic glasses, which are relatively new materials with limited data on structure and properties?  (2) How can we connect the timescale gap between MD simulations and experimental observations to gain accurate insights? (3) How can the ECNLE theory be refined to better capture the complex behavior of specific alloy systems? In this work, we address the above open questions by integrating MD simulations and ECNLE theory to study the glass transition and relaxation dynamics of CoCrNi alloys, \ce{Zr_{46}Cu_{46}Al_8}, \ce{Zr_{50}Cu_{40}Al_{10}}, and \ce{Zr_{64.13}Cu_{15.75}Ni_{10.12}Al_{10}}. Our approach provides new insights into the dynamic arrest and relaxation processes in these complex materials, enhancing the applicability of the ECNLE theory to predict glassy dynamics in mid-, high-entropy, and metallic glass alloys. 

\section{Theoretical background of The ECNLE theory}
We begin by briefly reviewing the foundational principles of the ECNLE theory \cite{1,2,3,4,5,6,7,8,9,30,39,40,41,42}. This theory models glass-forming liquids as hard-sphere fluids, which are characterized by particle density, $\rho$, and particle diameter, $d$. The static structure of these fluids quantified by the static structure factor, $S(q)$, and the radial distribution function (RDF), $g(r)$, is determined using the Percus-Yevick theory \cite{10}. $S(q)$ and $g(r)$ are both dependent on the particle density, which is directly related to the volume fraction in a fluid.

The motion of a single particle within this fluid is governed by interactions with neighboring particles, random noise, and frictional forces. These combined effects are described by a nonlinear stochastic equation. By solving this equation associated with the force-force correlation function, we can derive the free dynamic energy to describe local particle interactions. The analytical expression for this dynamic free energy is
\begin{eqnarray}
\frac{F_{dyn}(r)}{k_BT} &=& -3\ln\frac{r}{d}
\\ &-&\int_0^{\infty} dq\frac{ q^2d^3 \left[S(q)-1\right]^2}{12\pi\Phi\left[1+S(q)\right]}\exp\left[-\frac{q^2r^2(S(q)+1)}{6S(q)}\right],\nonumber
\label{eq:1}
\end{eqnarray}
where $k_B$ denotes the Boltzmann constant, $T$ is the ambient temperature, $r$ is the displacement, $q$ is the wavevector, and $\Phi=\rho\pi d^3/6$ is the volume fraction of the system. The second term accounts for the caging effect, where a particle's motion is restricted by its neighbors. The first term characterizes the behavior of an ideal fluid.

\begin{figure*}[htp]
\includegraphics[width=13.5cm]{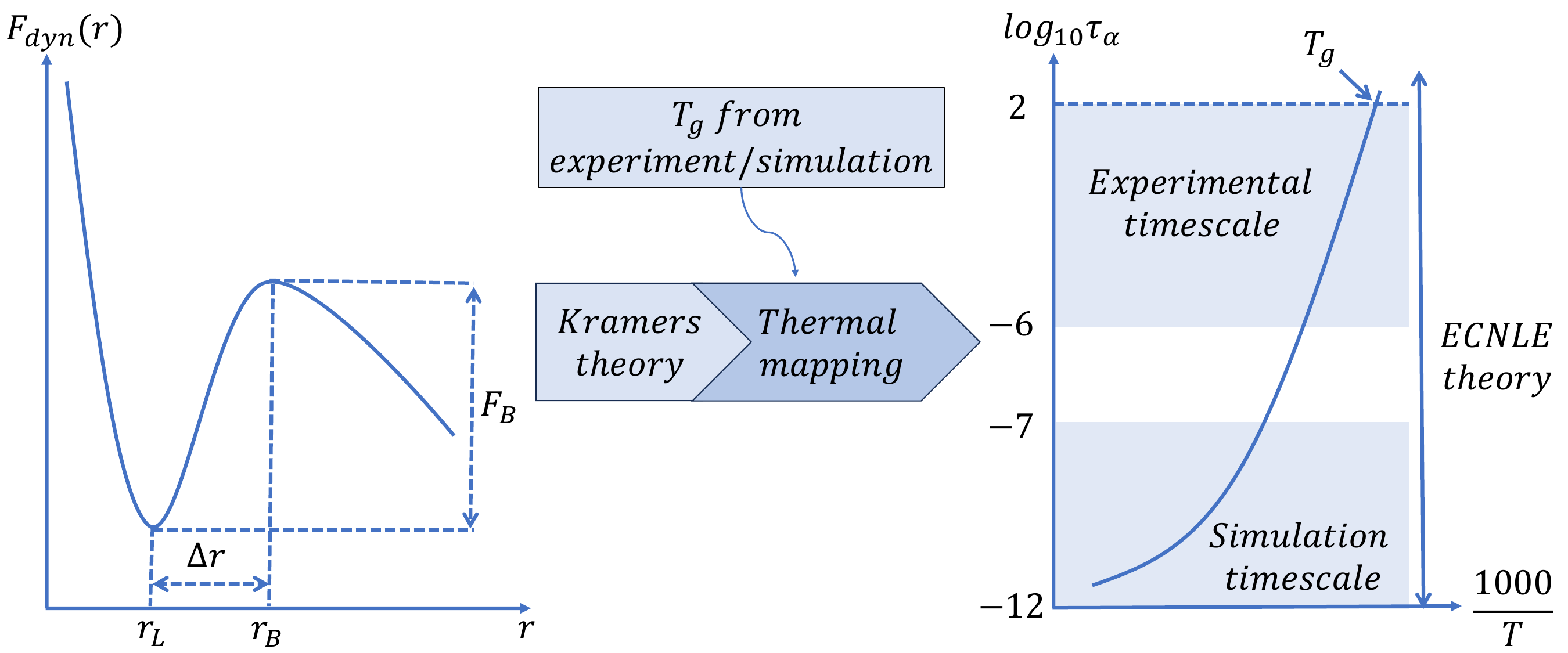}
\caption{\label{fig:0}(Color online) Illustration of the dynamic free energy profile for a tagged particle showing key length scales and the barrier height that governs local dynamics. These parameters are used in Kramers' theory and the thermal mapping to determine the structural relaxation time as a function of temperature.}
\end{figure*}

The free energy profile provides essential insights into the mechanisms governing local particle dynamics. At low densities ($\Phi < 0.432$), particles move freely as if in the fluid state due to weak interparticle interactions \cite{1,2,3,4,5,6,7,8,9,30,39,40,41}. In denser systems, interparticle interactions create a cage-like confinement around a central particle with the cage radius, $r_{cage}$, defined by the first minimum in the radial distribution function. Dynamic arrest occurs when a local energy barrier appears in the dynamic free energy landscape as depicted in Fig. \ref{fig:0}. This barrier height is $F_B=F_{dyn}(r_B)-F_{dyn}(r_L)$ with $r_L$ and $r_B$ being a local minimum and maximum of $F_{dyn}(r)$, respectively, and $\Delta r = r_B-r_L$ is the jump distance.

While long-range elastic correlations have been proposed as a potential contributor to caging effects in some systems \cite{1,2,3,4,5,6,7,8,9,39,41}, our previous work specifically on metallic glasses \cite{30,40} indicated that the influence of cooperative dynamics on the glass transition is relatively minor. Thus, in the present work, we consider only local cage effects to describe glassy dynamics in metallic glasses. The structural relaxation time as a function of volume fraction is determined using Kramers' theory, which gives
\begin{eqnarray}
\frac{\tau_\alpha}{\tau_s} = 1+ \frac{2\pi}{\sqrt{K_0K_B}}\frac{k_BT}{d^2}\exp\left(\frac{F_B}{k_BT} \right),
\label{eq:4}
\end{eqnarray}
where $\tau_s$ is a short time scale, $K_B$=$\left|\partial^2 F_{dyn}(r)/\partial r^2\right|_{r=r_B}$ is the absolute curvature at the barrier position, and $K_0 = \left|\partial^2 F_{dyn}(r)/\partial r^2\right|_{r=r_L}$ is a spring constant determined by the curvature of the dynamic free energy at the equilibrium distance $r_L$. $\tau_s$ is analytically obtained by \cite{1,2,3,4,5,6,7,8,9,30,39,40}
\begin{eqnarray}
\tau_s=g^2(d)\tau_E\left[1+\frac{1}{36\pi\Phi}\int_0^{\infty}dq\frac{q^2(S(q)-1)^2}{S(q)+b(q)} \right],
\label{eq:5}
\end{eqnarray}
where $\tau_E \approx 10^{-13}$ s is the characteristic Enskog time scale, $b(q)$ is determined by $b(q)=1/\left[1-j_0(q)+2j_2(q)\right]$, and $j_n(x)$ is the spherical Bessel function of order $n$ \cite{1,2,3,4,5,6,7,8,9,30,39,40}.

To quantitatively contrast our theoretical calculations against experimental/simulation data, it is necessary to relate the simulated packing fraction to a corresponding real-world temperature. In prior works \cite{1,2,3,4,5,30}, we proposed a thermal mapping model based principle of thermal expansion, which is
\begin{eqnarray}
T \approx T_g + \frac{\Phi_g - \Phi}{\beta\Phi_0},
\label{eq:7}
\end{eqnarray}
where $\beta = 12\times10^{-4}$ $K^{-1}$ is a typical volumetric thermal expansion coefficient for numerous amorphous materials, $\Phi_0 = 0.5$ is a characteristic packing fraction, the volume fraction $\Phi_g\approx 0.6715$ 
corresponds to the glass transition state of metallic glasses where the relaxation time $\tau_\alpha(\Phi_g)$ reaches 100 seconds, and $T_g$ is the glass transition temperature defined by $\tau_\alpha(T=T_g) = 100$ s. The $T_g$ value can be determined experimentally using techniques such as differential scanning calorimetry and broadband dielectric spectroscopy, or computationally through simulations. Previous works \cite{1,2,3,4,5,6,7,8,9,30,39,40,41,42} have successfully employed experimental $T_g$ values in ECNLE calculations (Eqs. (1-4)) to accurately predict the $\alpha$-relaxation time as a function of temperature for various amorphous materials. The predicted timescales span both experimental and simulation regimes as illustrated in Fig. \ref{fig:0}. However, the potential of MD-derived $T_g$ values for the ECNLE theory remains unexplored. In this study, we present the first application of the ECNLE theory using $T_g$ values obtained from MD simulations.

\section{RESULTS AND DISCUSSION}
In a prior work \cite{30}, we employed experimental $T_g$ values to predict the temperature dependence of structural relaxation times for several metallic glasses including \ce{Pd_{30}Ni_{50}P_{20}}, \ce{Pd_{40}Ni_{40}P_{20}}, \ce{Pd_{40}Ni_{10}Cu_{30}P_{20}}, \ce{Pd_{42.5}Ni_{7.5}Cu_{30}P_{20}}, and \ce{Zn_{38}Mg_{12}Ca_{32}Yb_{18}}. The theoretical predictions exhibited quantitative agreement with experimental data. For new metallic glasses, we determine $T_g$ by analyzing $g(r)$ obtained from MD simulations. The RDF provides insights into the spatial arrangement of atoms within the material. In multi-component systems, $g(r)$ consists of multiple partial contributions arising from distinct pairwise interactions between chemical elements. We compute and normalize the partial RDFs for all elemental pairs, weighting them by the product of their corresponding concentrations, to obtain the total RDF. The calculated total RDF is then utilized to determine the glass transition temperature in this study and compared with previous results. As a metallic glass is cooled, the RDF evolves and a distinct change emerges at the glass transition. Following the approach of Wendt and Abraham \cite{31}, by tracking the evolution of the ratio $g_{min}/g_{max}$ during cooling simulations with $g_{min}$ and $g_{max}$ being the first minimum and maximum of $g(r)$, respectively, the glass transition temperature can be precisely identified. Changes in alloy composition affect the partial radial distribution functions and their weights in the total RDF, and thereby influence the glass transition temperature.

\begin{figure*}[htp]
\includegraphics[width=5.9cm]{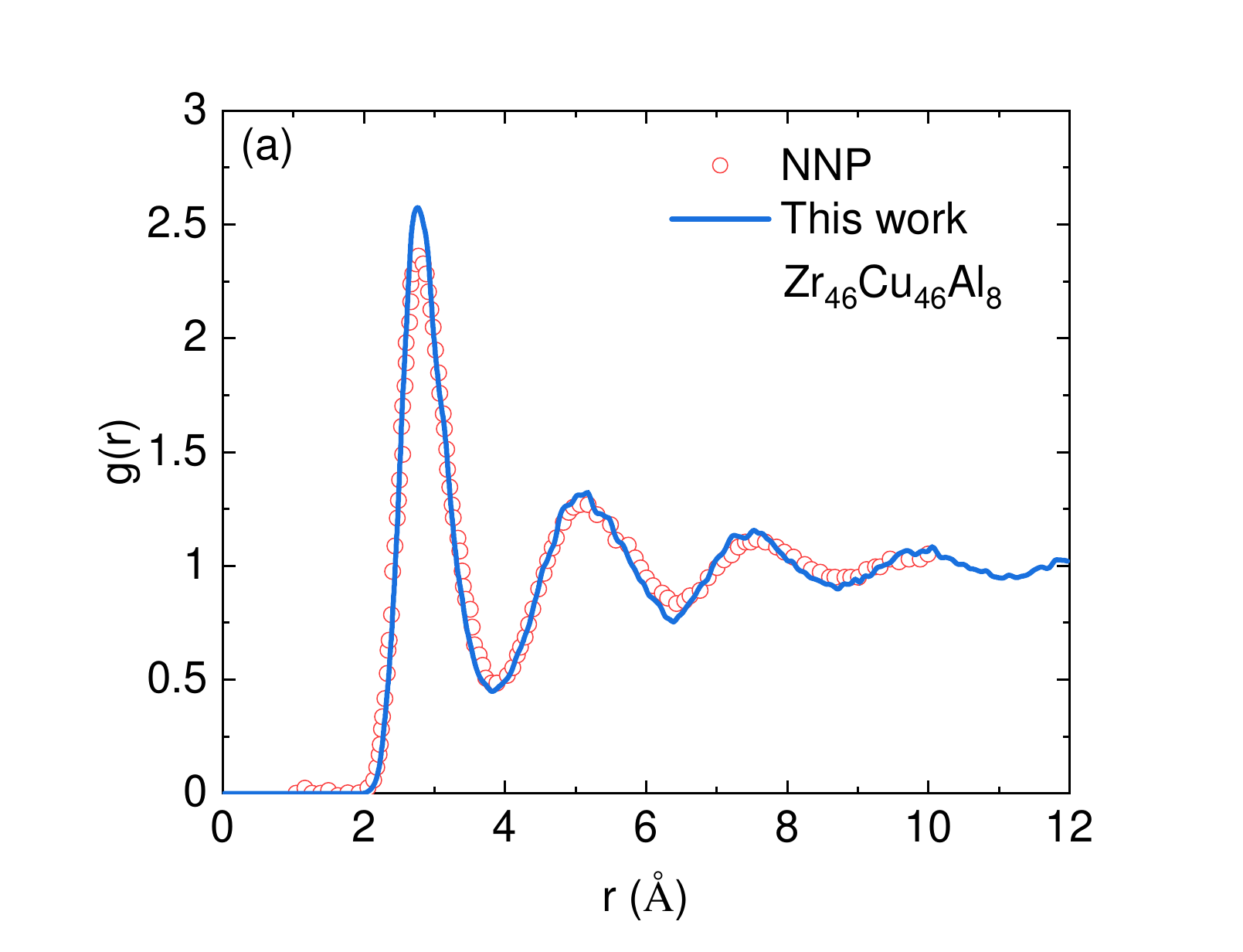}
\includegraphics[width=5.9cm]{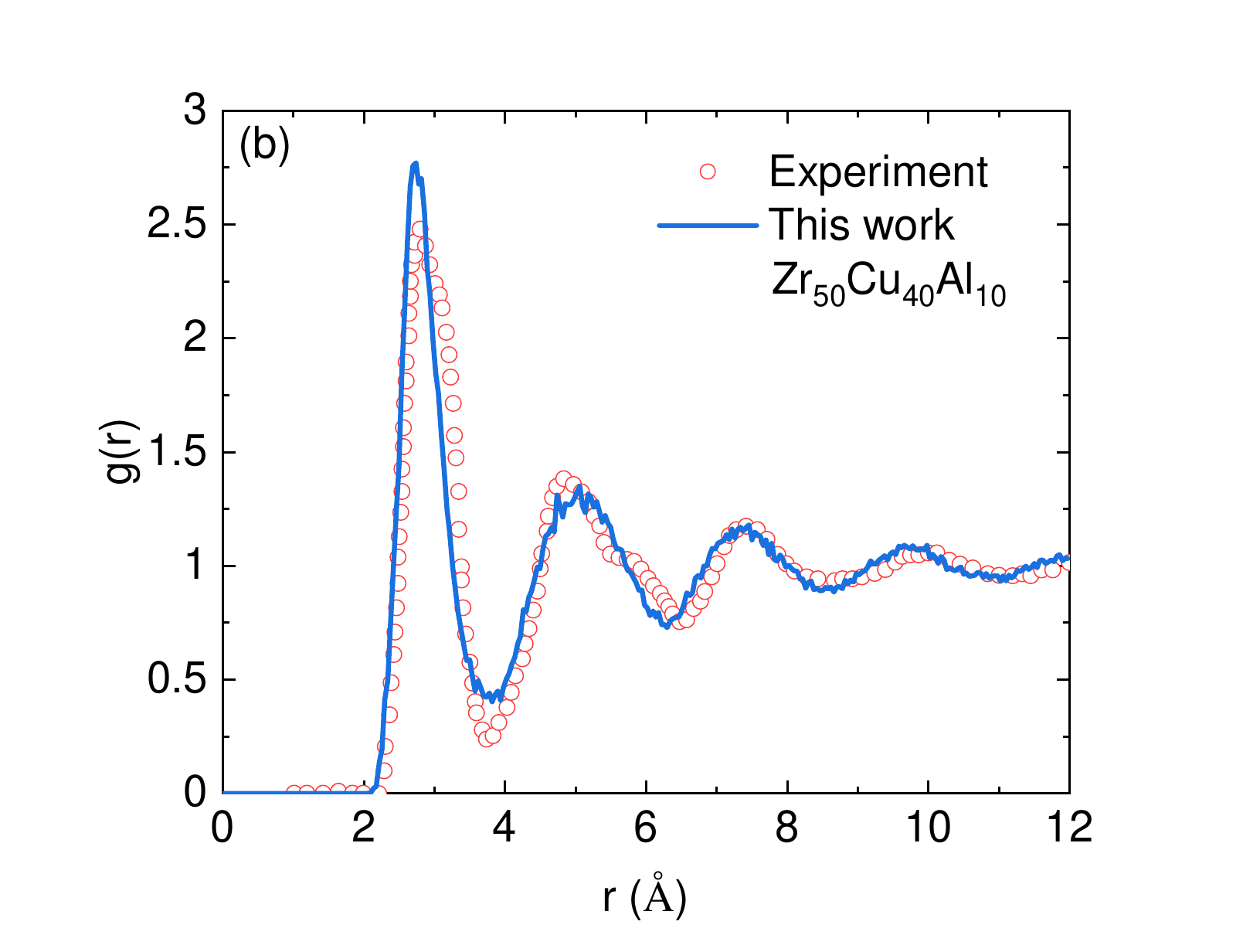}
\includegraphics[width=5.9cm]{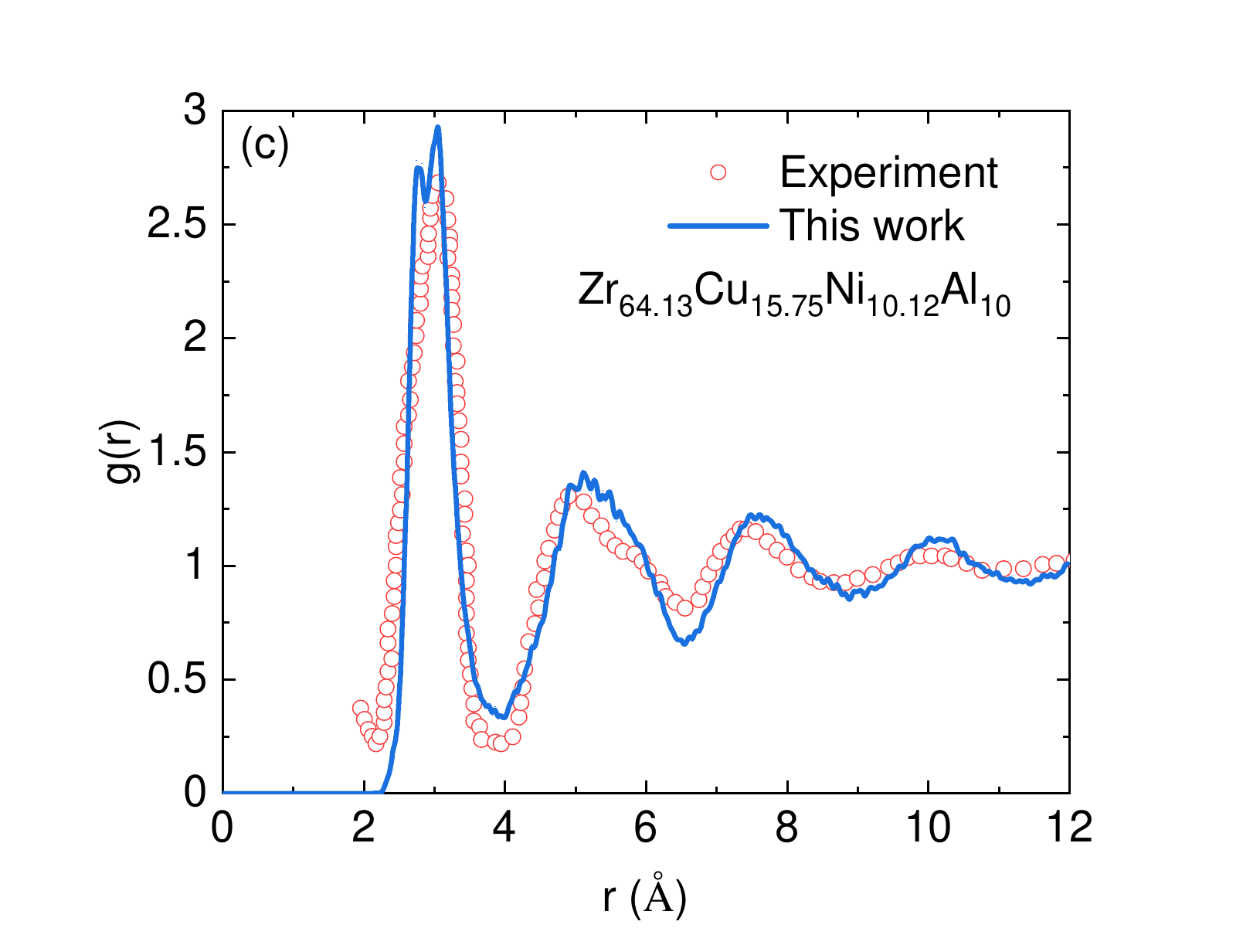}
\caption{\label{fig:1}(Color online) Our predicted RDFs and their counterparts for (a) \ce{Zr_{46}Cu_{46}Al_8} predicted using neural network potentials (NNP) at 1100 K \cite{15}, (b) \ce{Zr_{50}Cu_{40}Al_{10}} \cite{20}, and (c) \ce{Zr_{64.13}Cu_{15.75}Ni_{10.12}Al_{10}} \cite{24} measured by the synchrotron X-ray diffraction at 870 K and 298 K, respectively.}
\end{figure*}

All MD simulations were performed using the Large-scale Atomic/Molecular Massively Parallel Simulator (LAMMPS). Each simulation cell contains 4000 atoms interacting with each other via embedded-atom potentials \cite{32,33}. The initial systems were constructed as solid solutions with fcc crystal structures, where lattice sites were randomly occupied by the constituent elements. These systems were then melted and equilibrated at 3000 K for 40 ps under NPT conditions, using a 1 fs time step. Subsequently, the melts were gradually quenched to approximately 300 K over a timescale of 250 ps. The procedure is followed by an additional 40 ps of relaxation at each temperature. Periodic boundary conditions were applied in all three directions.

Before determining $T_g$ for \ce{Zr_{46}Cu_{46}Al_8}, \ce{Zr_{50}Cu_{40}Al_{10}}, and \ce{Zr_{64.13}Cu_{15.75}Ni_{10.12}Al_{10}}, we first validate our MD simulations by comparing the RDFs at fixed temperatures with experimental data and previous simulation studies \cite{15,20,24}. Figure \ref{fig:1} presents the quantitative agreement between our results and these reference works. Notably, the authors in Ref. \cite{15} performed density functional theory (DFT) calculations alongside neural-network-potential MD (NNP-MD) simulations for \ce{Zr_{46}Cu_{46}Al_8}. They showed an almost perfect overlap between the DFT and NNP-MD curves. This high level of agreement indicates that our results are also consistent with DFT simulations. Furthermore, despite the inherent complexities of experimentally fabricated metallic glass structures, the strong correlation between our simulation data and in-situ X-ray structural measurements for the metallic glasses \ce{Zr_{50}Cu_{40}Al_{10}} and \ce{Zr_{64.13}Cu_{15.75}Ni_{10.12}Al_{10}} shows the reliability and accuracy of our simulation approach.

\begin{figure}[htp]
\includegraphics[width=8.5cm]{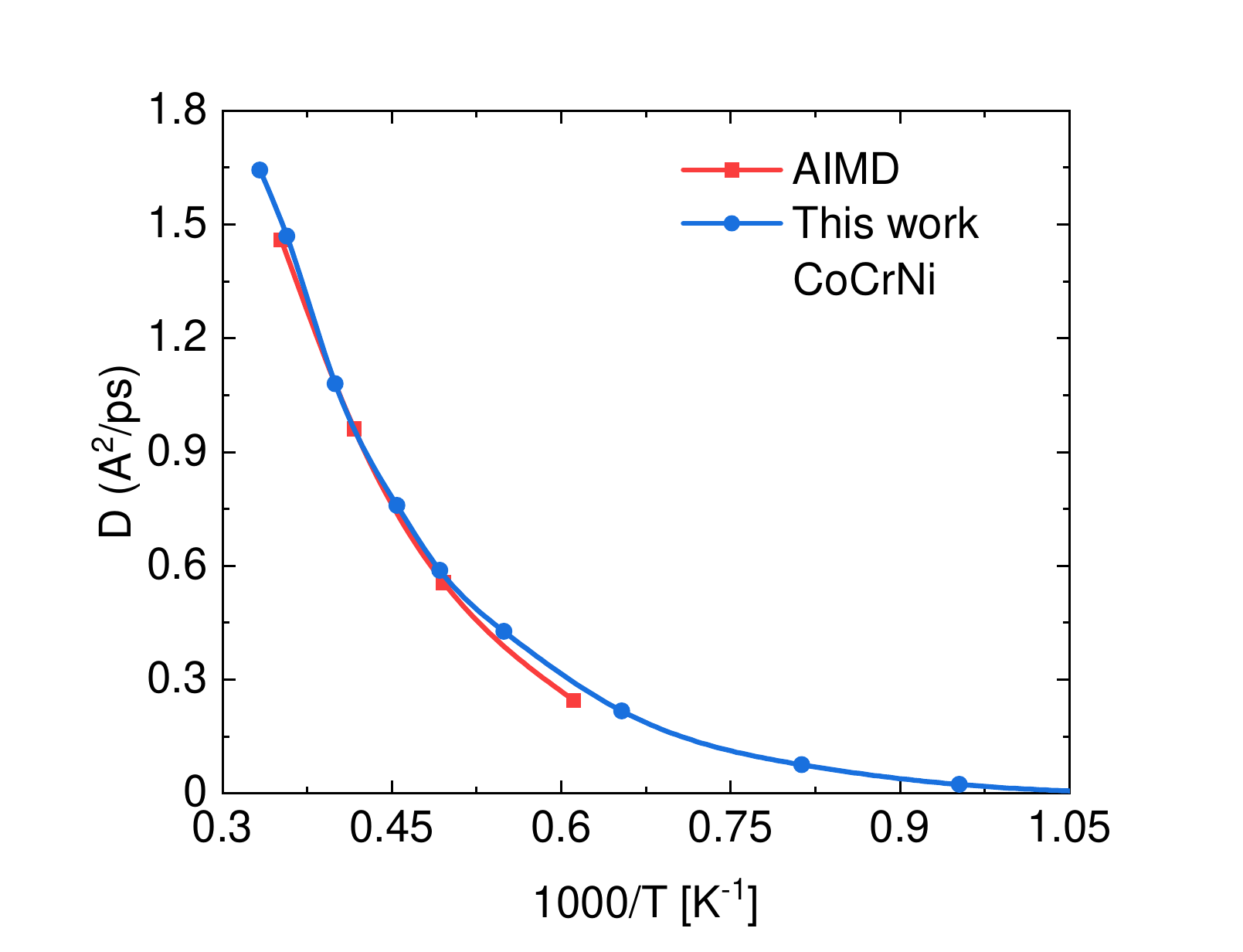}
\caption{\label{fig:2}(Color online) The concentration-weighted average of the self-diffusion coefficient for CoCrNi as a function of temperature computed in our work and by AIMD simulation in Ref. \cite{25}.}
\end{figure}

For the CoCrNi metallic glass, we utilize MD simulations to calculate the average diffusion coefficient at various temperatures. The diffusion coefficient, $D$, is obtained by analyzing the mean-square displacement (MSD) of atoms through the Einstein relation \cite{37}
\begin{eqnarray}
D = \lim_{t\to\infty}\left(\frac{1}{N}\sum_{i=1}^N \frac{1}{6t}\langle |\mathbf{r}_i(t) - \mathbf{r}_i(0)|^2 \rangle\right)
\label{eq:8}
\end{eqnarray}
where $\langle \cdot \rangle$ denotes an ensemble average over all particles $i$th in the system, $\mathbf{r}_i(t)$ is the position of particle $i$ at time $t$, $\mathbf{r}_i(0)$ is its initial position, and $N$ is the number of atoms. As illustrated in Fig. \ref{fig:2}, our numerical results show excellent quantitative agreement with diffusion coefficients obtained from \emph{ab initio} molecular dynamics (AIMD) simulations reported in Ref. \cite{15}. Despite the differences in methodology between MD and AIMD, the diffusion coefficients demonstrate remarkable consistency. Since the diffusion coefficient is directly derived from the atomic positions within the alloy over defined time intervals, this high level of accuracy reinforces the validity of our simulated RDFs.

\begin{table}[htp]
\caption{The glass transition temperatures in Kelvin computed in our work and their corresponding experimental and simulation values in previous works.}
\centering 
\begin{tabular}{|c | c| c|} 
\hline
Material & $T_g$ (this work) & $T_g$ (prior works)\\ [0.5ex] 
\hline 
CoCrNi & 750 & \\
[0.5ex] 
\hline
\ce{Zr_{46}Cu_{46}Al_8}& 769 & 708 \cite{11,13}, 695 \cite{12}, \\
 &   & 715 \cite{14}, 706 \cite{15},\\
 &   & 765 \cite{16}\\
[0.5ex]
\hline
\ce{Zr_{50}Cu_{40}Al_{10}} &  756 & 692 \cite{17}, 706 \cite{18}, \\
 &   & 703 \cite{20}, 673-717 \cite{19}\\
[0.5ex]
\hline
\ce{Zr_{64.13}Cu_{15.75}Ni_{10.12}Al_{10}} & 735.5 & 639 \cite{22}, 675 \cite{23},\\
  &   & 625 \cite{24} \\
[0.5ex]
\hline
\end{tabular}
\label{table:1} 
\end{table}

After validating the accuracy of our MD simulations, we compute the RDFs for the metallic glasses under study at various temperatures to determine their glass transition temperatures. The MD-simulated $T_g$ values are then quantitatively compared with those reported in previous studies as shown in Table \ref{table:1} Differences between the $T_g$ values predicted by MD simulations and experimentally measured values can be attributed to two main factors. In MD simulations, the accessible timescales are on the order of nanoseconds, which is substantially shorter than those in experimental conditions. The glass formation in experiments occurs at much slower cooling rates spanning from minutes to hours. This discrepancy in cooling rates leads to higher $T_g$ values in simulations, as the rapid quenching rates can trap the atomic structure at higher temperatures than those observed experimentally. Additionally, MD simulations generally use simplified interatomic potentials and idealized conditions, which may not fully capture the complexities of real materials such as defects, impurities, or structural heterogeneities. These complexities also contribute to discrepancies in $T_g$. In our case, the accelerated cooling rate in MD appears to be the primary factor contributing to the difference. Nonetheless, the deviation between our simulated $T_g$ values and those reported in other studies is less than 50 K (approximately 7 $\%$), which is within an acceptable range for subsequent analysis.

\begin{figure}[htp]
\includegraphics[width=8.5cm]{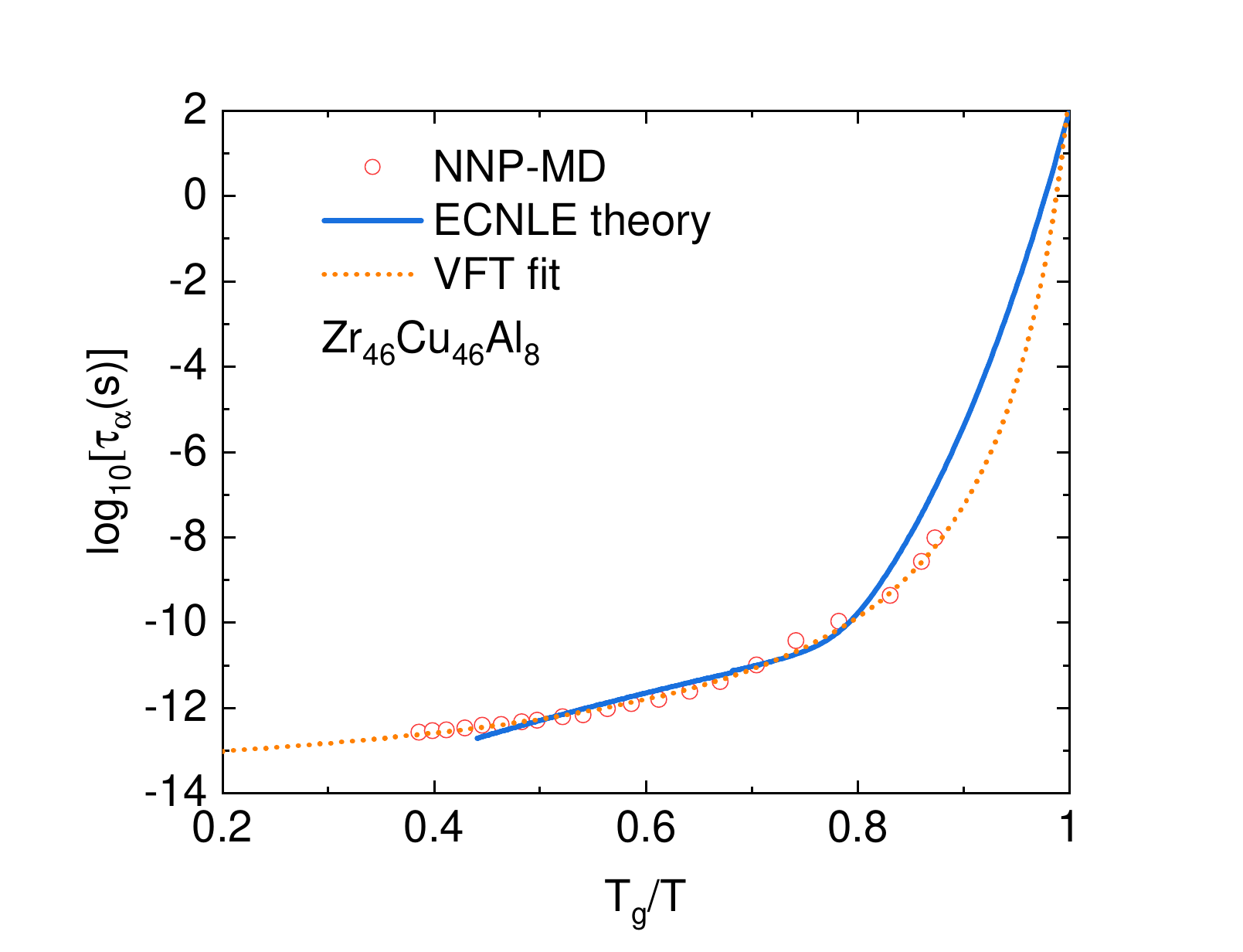}
\caption{\label{fig:3}(Color online) The logarithm of the structural relaxation time (in seconds) for the \ce{Zr_{46}Cu_{46}Al_{8}} metallic glass as a function of the inverse reduced temperature. The solid curve and data points correspond to ECNLE predictions and NNP-MD simulation data from Ref. \cite{15}. The dotted curve represents a VFT fit to the NNP-MD data. The VFT equation is $\log_{10}\tau_\alpha=\log_{10}\tau_\infty+B/(T-T_0)$ where the fitting parameters are $\log_{10}\tau_\infty = -13.277$, $T_0 = 265.3$ $K$, and $B=766.6$ $K$.}
\end{figure}

After determining $T_g$ from our MD simulation, we substitute this value into Eq. (\ref{eq:7}) along with Eqs. (\ref{eq:1}-\ref{eq:7}) to calculate the temperature dependence of the structural relaxation time by the ECNLE theory. For \ce{Zr_{46}Cu_{46}Al_{8}}, we obtained a $T_g$ of 750 $K$. Notably, as shown in Fig. \ref{fig:3}, our ECNLE predictions for $\tau_\alpha(T)$ demonstrate strong quantitative agreement with the NNP-MD simulation data reported in Ref. \cite{15}. Note that the timescale of NNP-MD simulations (less than 10 ns) is significantly shorter than the experimental relaxation timescale ($\sim$ 100 s). To facilitate comparison with experimental data or our ECNLE predictions on an experimental timescale, the VFT function was employed to extrapolate the NNP-MD data to longer timescales. While our ECNLE calculations show reasonable agreement with the extrapolated VFT results, the VFT extrapolation suggests a higher fragility compared to our theoretical predictions.

\begin{figure}[htp]
\includegraphics[width=8.5cm]{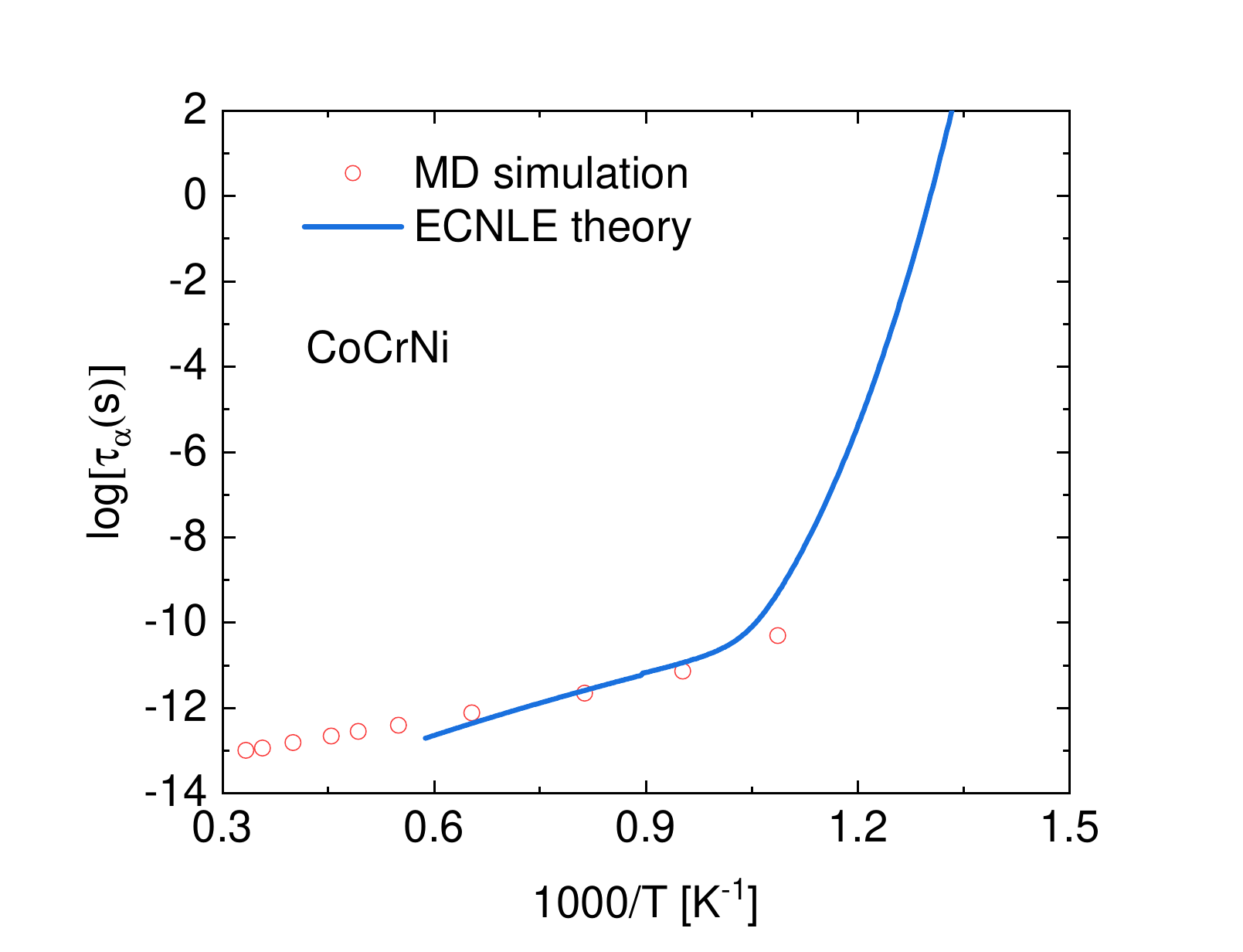}
\caption{\label{fig:4}(Color online) The logarithm of the structural relaxation time (in seconds) for the CoCrNi metallic glass as a function of $1000/T$ calculated using the ECNLE theory (solid curve) and MD simulation (data points).}
\end{figure}

For the CoCrNi mid-entropy alloy, our MD simulations predict $T_g=750$ $K$. This value combined with the ECNLE theory allows us to calculate the structural relaxation time $\tau_\alpha(T)$ as shown in Fig. \ref{fig:4}. While direct experimental or simulation data for $\tau_\alpha(T)$ of CoCrNi is unavailable,  we can infer it from the diffusion coefficient. According to the ECNLE theory, $D(T)$ and $\tau_\alpha(T)$ are related by 
\begin{eqnarray}
D(T) = \frac{\Delta r^2}{6\tau_\alpha(T)}.
\label{eq:9}
\end{eqnarray}
where $\Delta r\approx0.2-0.4d$ \cite{1,2,3,4,5,6,7,8,9,30}. In our ECNLE framework, the effective diameter of a hard-sphere particle corresponds to the average lattice spacing of the alloy’s unit cell. Based on our MD simulations, $d \approx 3.6$ $\AA$ and, thus, $\Delta r \approx 1$ $\AA$. By substituting these values of $\Delta r$ and the calculated $D(T)$ from Fig. \ref{fig:2} into Eq. (\ref{eq:9}), we obtain $\tau_\alpha(T)$. Interestingly, the ECNLE calculations quantitatively agree with our MD-derived estimates. The quantitative agreement between our ECNLE calculations and MD simulations for CoCrNi suggests that this approach can serve as a reliable tool for predicting structural relaxation times in a wide range of metallic glass systems.

As mentioned above, for $\Phi \leq0.432$, the system exhibits fluid-like behavior with delocalized particles, similar to a high-temperature melt. This suggests that the volume fraction $\Phi_l=0.432$ may correspond the liquidus point of the hard-sphere fluid as it represents the boundary between the normal liquid and supercooled regimes. Based on this assumption, Eq. (\ref{eq:7}) predicts a temperature difference of $T_l-T_g\approx 400$ $K$. This result leads to an estimated liquidus temperature for \ce{Zr_{46}Cu_{46}Al_8} of approximately 1168 K, which is quantitatively close to the reported value of 1163 $K$ in Ref. \cite{11} and the range of 1209-1229 $K$ in Ref. \cite{15}. Furthermore, our analysis of 715 metallic glasses from Ref. \cite{38} reveals that $T_l-T_g \approx 464 \pm 97$ $K$. This finding is also consistent with our result. Note that this work neglects the influence of elastic collective dynamics on the glass transition. Incorporating these collective effects, though small, decreases $\Phi_g$ \cite{3,5} and, thereby, narrows the gap between the liquidus and glass transition temperature. 

\section{Conclusions}
In conclusion, our study has presented a comprehensive investigation of the glass transition and structural relaxation dynamics in CoCrNi and other metallic glasses through MD simulations and the ECNLE theory. By reasonably determining the glass transition temperature from MD simulations and employing the ECNLE theory, we successfully predicted the temperature-dependent behavior of the structural relaxation time and diffusion constant. Numerical results quantitatively agree with experimental data and simulations in previous works \cite{15,20,24,25}. The quantitative agreement validates our approach to investigate relaxation dynamics in metallic glasses. This multiscale approach, which combines atomistic simulations with force-level statistical mechanics, allows us to provide a more profound understanding of relaxation dynamics and diffusion across different timescales. This work advances our understanding of the fundamental physics governing glass-forming alloys and paves the way for the design of materials with optimized physical-mechanical properties.

\begin{acknowledgments}
This research was funded by the Vietnam National Foundation for Science and Technology Development (NAFOSTED) under Grant No. 103.01-2023.62.
\end{acknowledgments}
\section*{Conflicts of interest}
The authors have no conflicts to disclose.

\section*{Data availability}

The data that support the findings of this study are available from the corresponding author upon reasonable request.

\end{document}